\documentclass[aps,prd,twocolumn,reprint,nofootinbib]{revtex4-1}
\usepackage{graphicx}  

\usepackage{dcolumn}  
\usepackage{bm}        
\usepackage{amssymb}
\usepackage{amsmath} 
\usepackage{epstopdf} 
\usepackage{amsfonts,float}
\usepackage{fancyhdr}
\usepackage{pifont}
\usepackage[hidelinks]{hyperref} 
\usepackage{bbm}
\usepackage{tensor}
\usepackage{mathrsfs}
\usepackage{xcolor}
\usepackage{physics}

\usepackage{color} 

\usepackage[titles]{tocloft}
\newcommand{\bea}{\begin{eqnarray}} 
\newcommand{\eea}{\end{eqnarray}}
\newcommand{\beq}{\begin{equation}} 
\newcommand{\eeq}{\end{equation}}

\newcommand{\qed}{\nobreak \ifvmode \relax \else
      \ifdim\lastskip<1.5em \hskip-\lastskip
      \hskip1.5em plus0em minus0.5em \fi \nobreak
      \vrule height0.75em width0.5em depth0.25em\fi}

\begin{document}

\title{Casimir energy for concentric $\delta$-$\delta'$ spheres}
\author{In\'es Cavero-Pel\'aez}\email{cavero@unizar.es.com}\affiliation{Centro Universitario de la Defensa,
  Zaragoza, 50019, Spain. \\  Departamento de F\'isica Te\'orica, Facultad de Ciencias,\\ Universidad de Zaragoza, Zaragoza, 50009, Spain.}\author{J. M. Munoz-Castaneda}\email{jose.munoz.castaneda@uva.es}\affiliation{Departamento de F\'isica Te\'orica, At\'omica y Optica,Universidad de Valladolid,
  Valladolid, 47011, Spain.} \author{C. Romaniega}\email{cesar.romaniega@uva.es}\affiliation{Departamento de F\'isica Te\'orica, At\'omica y Optica,Universidad de Valladolid,
  Valladolid, 47011, Spain.}

\begin{abstract}

We study the vacuum interaction of a scalar field and two concentric spheres defined by a singular potential on their surfaces. The potential is a linear combination of the Dirac $\delta$ and its derivative. The presence of the delta prime term in the potential causes that it behaves differently when it is seen from the inside or from the outside of the sphere. We study different cases for positive and negative values of the delta prime coupling, keeping positive the coupling of the delta. As a consequence, we find regions in the space of couplings, where the energy is positive, negative or zero. Moreover, the sign of the $\delta'$ couplings cause different behaviour on the value of the Casimir energy for different values of the radii. This potential gives rise to general boundary conditions with limiting cases defining Dirichlet and Robin boundary conditions what allows us to simulate purely electric o purely magnetic spheres. 
\end{abstract}
 \maketitle

\section{Introduction}
Casimir forces are measurable effects arising when vacuum fluctuations of quantum fields are modified by external conditions such as bodies with different geometries or boundaries. Among many others, some examples can be found in Refs. \cite{klimchitskaya2009casimir,munday2009measured,chan2001quantum}. A large amount of studies for different geometries  have been carried out over the years, where a great deal of the work has focused on the interaction energy between bodies \cite{bordag2009advances,dalvit2011casimir,milton2004casimir}. This makes sense since it is feasible to setup an experiment that measures forces between objects. The interpretation of the Casimir interaction energy is clearer and less controversial than that of the Casimir self-energy of a single body, where surface divergences are still an open subject \cite{candelas1982vacuum,candelas1986vacuum,cavero2006local}. As in the original setup proposed by Casimir \cite{casimir1948attraction}, most of the systems studied present two objects  outside each other, even though  other configurations like cavities are experimentally realizable. In this context, a lot of work has been focused on systems as long cylinders \cite{mazzitelli2003casimir}, configurations of spheres \cite{brevik2002casimir,saharian2001scalar,teo2012mode} or Casimir-Polder interactions with a polarizable particle \cite{marachevsky2001casimir,milton2011casimir}.\\

In the line with what is mentioned above, the systems studied with the practical formulation based on functional determinants proposed in 2006 by Kenneth and Klich \cite{kenneth2006opposites,emig2008casimir,kenneth2008casimir} also focus on separated interacting bodies.  Even though there is no restriction on the disposition of the objects as long as they do not overlap, most of the attention has been directed towards bodies outside each other. However, if one body is inside the other, the socalled TGTG formula is still valid, although interior and exterior scattering must be considered. To our knowledge, this was firstly  discussed in Ref.~\cite{rahi2009scattering} for the electromagnetic field.  With this formalism new results were obtained: corrections to the Proximity Force Approximation (PFA) \cite{zaheer2010casimir}, analysis of the torque and alignment of a spheroid inside a cavity \cite{rahi2010stable} and the stability of certain collections of objects \cite{rahi2010constraints}, where some previous plausible configurations for stable levitation were discarded. In Ref.~\cite{teo2012mode} it is studied the scalar and electromagnetic fields interacting in the presence of two bodies outside each other or one inside the other whose centers are separated a certain distance. Bimonte \cite{bimonte2016} particularized the formula to the interaction between two perfectly conducting spheres in different arrangements, including the concentric case while in \cite{bordag2009vacuum} Bordag et al. calculate the interaction energy of a cylinder inside another beyond PFA, reaching the next-to-leading order correction to the vacuum energy for some cases of the radii of the cylinders.\\

In this work we compute the interaction energy  between two partially transparent concentric spheres for a  scalar field using the TGTG formula. Quantum vacuum energies for spherical backgrounds have been largely studied (see e.g. \cite{beauregard2015casimir,BordagPRD1996}). When the potentials mimicking the spherical shells do not overlap the total quantum vacuum energy of the system is given by
\begin{equation}
E_0=E(S_1)+E(S_2)+E_C,
\end{equation}
where the first two terms are the self-energies of the spheres individually and $E_C$ is the interaction term that we are interested in. The self-energies contain a finite and a divergent term: $E(S_i)=E^\text{fin}(S_i)+E^\text{div}(S_i)$. For massive fields there is a well defined unique finite part of the self-energy after imposing the mass renormalization condition and the divergences have been computed analytically \cite{beauregard2015casimir}. 
For the massless case a well defined result regularization-independent depends on the cancellation of certain heat kernel coefficients, specifically the coefficient $a_2$. When this cancellation does not occur, the finite part of the self-energy depends on a regularization parameter that needs a physical interpretation \cite{Fulling89}. This problem will be addressed in a forthcoming publication. Since $E(S_i)$ depends on the radius of the sphere, it contributes to the total quantum vacuum pressure over the spherical shell $S_i$\footnote{As a consequence, the main statement in Ref. \cite{CRS20} refers only to the contribution to the total pressure arising from the interaction terms of the energy. Nevertheless the self-energy contribution has been considered for a particular case, obtaining an expansion tendency ($p>0$).}. However, in the present paper we focus our attention on the part of the quantum vacuum energy that couples both spherical shells , i.e. $E_C$, which we call quantum vacuum interaction energy.
\\

The properties of each sphere enter in the interaction energy only through its $\mathbb{T}$ matrix \cite{lippmann1950variational}, which can be  easily calculated for spherical bodies. We mimic the spheres by a generalization of the  Dirac $\delta$  spherical shell, the socalled $\delta$-$\delta'$ interaction \cite{munoz2019hyperspherical}. Configurations based on the  $\delta$ potential  have widely  appeared in the literature, just to name some: $\delta$ sphere in two and three dimensions \cite{bordag1999ground,beauregard2015casimir,barton2004casimir,cavero2007local,parashar2017electromagnetic},  concentric and nonconcentric $\delta$ spheres \cite{teo2012mode} for both scalar and electromagnetic fields and the interaction between two $\delta$ lattices \cite{bordag2015dirac,bordag2017casimir,bordag2014monoatomically}. The addition of the $\delta'$ term to the potential that defines the plates was firstly considered in Ref. \cite{munoz2015delta} in the context of  Casimir physics. This is useful, essentially, in two aspects. Firstly,  Robin  boundary conditions can be obtained as a finite limit as was shown in Ref. \cite{munoz2015delta}. Secondly, although it is still not well understood, the sign of the  force depends strongly on the boundary, switching from attractive to repulsive forces \cite{bordag2001new}. Basically, the only general result concerning this issue is  restricted to  mirror symmetric bodies, originally proposed in Ref.~\cite{kenneth2006opposites} and extended in Ref.~\cite{bachas2007comment}. As we shall prove, we can gain insight in the latter with this interaction. We are able to identify  the configurations  in which the energy is positive or negative as a function of the parameters that define the potential on the spheres. For parallel plates, this has already been proved in Ref. \cite{munoz2015delta}. \\

Specifically, the semitransparent $\delta$-$\delta'$ spheres will be defined by the potentials
\begin{equation}\label{eq:V}
\bar V_{i}(r)= a_{i} \hspace{0mm}\delta  (r-r_{i})+b_{i}\hspace{0mm}\delta' (r-r_{i}),\quad a_{i},b_{i}\in \mathbb{R}\quad i=1, 2, 
\end{equation}
where $r_1$ and $r_2$ are the  radii of the inner and outer sphere ($r_1<r_2$), respectively. The definition of the previous potential is given by suitable matching conditions imposed on the scalar field \cite{munoz2019hyperspherical}. These conditions come from the  original work of Kurasov \cite{kurasov1996distribution} in one dimensional systems. The main advantage of these singular potentials is that they are often exactly solvable and therefore, provide a good
insight for some of the relevant quantum properties\footnote{ Despite its apparent simplicity,  there is a collection of applications in  a variety of areas in modern physics, see Ref.~\cite{romaniega2020approximation} and references quoted therein.}. 
Given the above,  the action that governs the dynamics of the massless scalar field interacting with this background is 
\begin{equation}\label{eq:Action}
S(\varphi)=\int d^{3+1}y\left[(\partial\varphi)^2-V(x)\varphi^2\right],
\end{equation}
where
\begin{eqnarray}\label{eq:Vx}
V(x)&=&V_1(x)+V_2(x)\nonumber \\
&=&\sum_{i=1}^{2} \lambda_{0,{i}} \hspace{0mm}\delta  (x-x_{i})+2 \lambda_{1,{i}}\hspace{0mm}\delta' (x-x_{i}).\label{eq:Vx}
\end{eqnarray}
We have chosen units such that $\hbar=c=1$ and introduced a mass parameter $\mu$ in order to work with dimensionless quantities,
\begin{eqnarray}
&&x \equiv r\mu,\quad x_{i} \equiv r_{i}\mu,\quad \varphi \equiv \frac{\phi}{\mu},\\
&& \lambda_{0,{i}}\equiv\frac{a_{i}}{\mu},\quad \lambda_{1,{i}}\equiv \frac{b_{i}}{2}.
\end{eqnarray} 
The paper is organized as follows. In section 2 we give an interpretation of the TGTG formula based on the mode summation approach when applied to the case of concentric spheres. In section 3 we study the solutions of the field modes for the potential under consideration and calculate the relevant elements of the TGTG formula, which allow us to give a simple expression for the interaction energy shown in section 4 together with some numerical results. Finally, we discuss these results and compare them with limiting cases. We finish in section 5 with the conclusions.  
%
\section{Scattering formalism interpretation}
The scattering approach to the computation of Casimir interaction energies between two bodies has been used in many calculations since more than {half a century}. It is worth to mention the original work of Balian and Duplantier in the 1970s \cite{balian1977electromagnetic,balian1978electromagnetic}. Most modern forms of calculating these energies have been developed by other authors already mentioned \cite{kenneth2006opposites,kenneth2008casimir,rahi2009scattering,graham2002calculating,jaffe2005casimir}. Their method has become very popular since it is free of divergences and it allows to obtain numerical results in a simple way. Examples of that are the interaction between a compact object and a plane \cite{emig2008fluctuation} or more specifically, between a sphere and plane \cite{canaguier2009casimir}. In those cases, the authors calculate the Casimir energy by computing the transition matrices of the scattered waves on the objects separately (the Lippmann-Schwinger operators of the bodies \cite{taylor2006scattering}), and the translation matrices from one object's origin to the other describing the propagation of the wave between them. In particular if we denote by $\mathbb T_{i}$, $i=1,2$, the Lippmann-Schwinger operators for each body, and $\mathbb U_{i,j}$, $i,j=1,2$, the free Green function that represents the translation from the center of body $i$ to the centre of body $j$ the socalled TGTG formula is given by
\begin{equation}\label{eq:EnergyTGTG}
  E_{\text{C}}=\frac{1}{2\pi}\int_0^\infty d\chi {\rm Tr}\,\ln (\mathbb I-\mathbb T_1\mathbb U_{12}\mathbb T_2\mathbb U_{21}).
\end{equation}
Here the integration is over the imaginary frequency.  Concerning our system of two concentric spheres, in Eq.~ \eqref{eq:EnergyTGTG} we denote with subindex 1, quantities referred to the interior sphere, and the subindex 2 refers to the analogues for the exterior sphere.

As we have previously stated, our case corresponds to the interaction of a scalar field on a background of concentric spheres with singular potential on their surfaces given by Eq.~\eqref{eq:V}. The great advantage of using Eq.~ \eqref{eq:EnergyTGTG} is that we only need to have information about each of the bodies individually. It is sufficient to know the shape of the incident and scattered waves and how they scatter on their surfaces, something that is determined by the boundary conditions on the spheres. They select the quantum fluctuations that will produce the interaction between the spheres. Due to the spherical symmetry of the problem, it is convenient to use spherical coordinates so that we expand the waves in the spherical harmonics from their origins. Since they share origin, the transition matrices become diagonal identities \cite{bimonte2016} (maybe multiplied by a constant depending on the normalization used).\\

The components of the Lippmann-Schwinger operators $\mathbb T_{1,2}$ of each object appearing in the TGTG formula that describes our system combine in such a way that the formula will pick up the
quantum vacuum fluctuations that give rise to the interaction. Since we have one body inside the other, it represents the scattering  produced by the exterior and interior sides of the spheres respectively. It is relevant to notice that the inner and the outer side of the $\delta$-$\delta'$ sphere do not produce the same interaction.
Therefore, each object will contribute with different components of the $\mathbb T$ operator. 
%
\section{General scattering solutions and $\boldsymbol{T}$ operators}
%
Let's consider now a single sphere defined by the potential given in section I,
\begin{equation}\label{eq:potU}
V_0(x)=\lambda_{0} \hspace{0mm}\delta  (x-x_0)+2 \lambda_{1}\hspace{0mm}\delta' (x-x_0),\quad x_0\in\mathbb{R}^+.
\end{equation}
Infinitesimal variations of the action in \eqref{eq:Action} impose that the scalar field $\varphi(t,\mathbf{x})$ satisfies the equation of motion
\begin{equation}
  - \partial_\mu\partial^\mu \varphi(t,\mathbf{x}) - V_0(x)\varphi(t,\mathbf{x})=0,
\end{equation}
where $\mu$ is an index that can take the values $\{0,1,2,3\}$. 
Since the potential is time independent, the Fourier transform in time allows us to work at a given frequency that later on, we integrate over the whole range. Then,
\begin{equation}
  \varphi(t,\mathbf{x})=\int_{-\infty}^{\infty}d\omega\, \varphi_\omega(\mathbf{x})e^{-i\omega t},\qquad \mathbf{x}\equiv(x, \theta, \phi). \nonumber
\end{equation}
The resulting equation can now be written as
\begin{equation}\label{eq:schrH}
  \left[-\Delta+V_0(x)\right]\varphi_\omega(\mathbf{x}) = \omega^2 \varphi_\omega(\mathbf{x}),
\end{equation}
where $\Delta$ is the Laplacian operator. The nonrelativistic Schr\"odinger Hamiltonian in Eq.~\eqref{eq:schrH} has been recently studied in detail in \cite{munoz2019hyperspherical}, where the potential $V_0(x)$ is defined by matching conditions at the sphere of dimensionless radius $x=x_0$ over the space of field modes as  
\begin{equation}\label{eq:matchG}
 \left(
    \begin{array}{c}
     \varphi(x_0^+,\theta,\phi) \\
      \dot\varphi(x_0^+,\theta,\phi) \\
    \end{array}
    \right)=\left(
    \begin{array}{cc}
      \alpha  & 0 \\
      \widetilde{\beta} & {\alpha^{-1} } \\
    \end{array}
    \right)\left(
    \begin{array}{c}
     \varphi(x_0^-,\theta,\phi) \\
      \dot\varphi(x_0^-,\theta,\phi) \\
    \end{array}
    \right),
\end{equation}
where we have introduced the notation
\begin{equation*}
\dot\varphi(\mathbf{x})\equiv\frac{\partial\varphi}{\partial x},
\end{equation*}
 $x_0^+$ and $x_0^-$ denotes that we approach $x_0$ from the right or from the left respectively and 
\begin{equation}\label{eq:defs}
\alpha\equiv \frac{1+\lambda_1}{1-\lambda_1},\quad\widetilde\beta\equiv \frac{\widetilde \lambda_{0}}{1- \lambda_{1}^2},\quad \widetilde \lambda_{0}\equiv -\frac{4\hspace{0mm} \lambda_{1}}{x_0}+{\lambda_{0}}.
\end{equation}
Due to the spherical symmetry of the potential, in Eq.~\eqref{eq:schrH} we  perform separation of variables that enables to expand the solution in the spherical harmonics $Y_{\ell m}(\theta,\phi)$ and write the modes of the field as 
\begin{equation}\label{solution_decomposition}
\varphi_\omega(\mathbf{x}) =\sum_{\ell=0}^\infty \sum_{m=-\ell}^\ell \rho_\ell(x)Y_{\ell m}(\theta,\phi).
\end{equation}
Accordingly, the radial modes $\rho_\ell(x)$ satisfy the differential equation
\begin{equation}\label{eq:Radial}
\left[- \frac{d^2}{dx^2}\! -\!\frac{2}{x}\frac{d}{dx}
+ \frac{\ell(\ell+1)}{x^2}-\omega^2\right]\rho_\ell(x)=0.
\end{equation}
Two independent solutions are $\rho_\ell^{{\rm reg}}(x)=j_\ell(\omega x)$  and $\rho_\ell^{{\rm out}}(x)=h_\ell^{(1)}(\omega x)$. The former is the spherical Bessel function regular at the origin and the latter is the spherical Hankel function of the first kind, which determines the radial part of a purely outgoing wave \cite{olver2010nist}.  Now we make the scalar field scatters with the sphere in two different situations.
\paragraph{The exterior scattering.} As mentioned above, the interior sphere of our system enters the TGTG formula through the component of the Lippmann-Schwinger operator that represents the scattering problem with the source and detector outside the object. In this sense, the general solution of a radial mode in the two regions separated by the sphere of radius $x_0$ is
\begin{equation}\label{solution_source_outside}
\rho_\ell(x)=\left\{ 
\begin{array}{cc}
A_\ell\rho_\ell^{{\rm reg}}(x) & x<x_0 \\
a_\ell\rho_\ell^{{\rm reg}}(x)+b_\ell\rho_\ell^{{\rm out}}(x)  & x>x_0  \\
\end{array}
\right. .
\end{equation}
If we impose matching conditions, given by Eq.~\eqref{eq:matchG}, on the surface of the sphere we obtain the system of equations,
  \begin{eqnarray}
   && \left(
    \begin{array}{c}
      a_\ell \rho_\ell^{\text{reg}}(x_0)+b_\ell \rho_\ell^{\text{out}}(x_0) \\
      { a_\ell \dot{\rho}_\ell^{\text{reg}}(x_0)+b_\ell \dot{\rho}_\ell^{\text{out}}(x_0)} \\
    \end{array}
    \right)\nonumber \\ 
    &=&A_\ell \left(
    \begin{array}{cc}
      \alpha  & 0 \\
      \widetilde{\beta} & {\alpha^{-1} } \\
    \end{array}
    \right)\left(
    \begin{array}{c}
      \rho_\ell^{\text{reg}}(x_0) \\
      \dot\rho_\ell^{\text{reg}}(x_0) \\
    \end{array}
    \right),\label{bc_outer_scattering}
  \end{eqnarray}
  where we have used the quantities and notation defined in Eq.~\eqref{eq:defs}. 
  Then the scattering produced by the sphere in this situation can be calculated as $$\mathbb T_{i}^\ell =-\frac{b_\ell}{a_\ell}.$$ Eliminating $A_\ell$ from Eq.~\eqref{bc_outer_scattering} we find
  \begin{eqnarray}
   \mathbb T^\ell(\omega)&=&\frac{j_\ell(\omega x_0)}{\Lambda(\omega)}\Big\{\big[\ell(\alpha^2-1)-x_0\alpha\tilde\beta\big]j_\ell(\omega x_0)\nonumber\\ && -(\alpha^2-1)\omega x_0\,j_{\ell+1}(\omega x_0)\Big\},\label{T_i}
  \end{eqnarray}
where
\begin{eqnarray}
    &&\Lambda(\omega)\equiv j_\ell(\omega x_0)\big[(\ell(\alpha^2-1)-\alpha\tilde\beta\,x_0)h_\ell^{(1)}(\omega x_0)\nonumber\\
    &&-\alpha^2\omega x_0 h_{\ell+1}(\omega x_0)\big]
    +\omega x_0 j_{\ell+1}(\omega x_0)h_\ell(\omega x_0).
\end{eqnarray}
    
\paragraph{The interior scattering.} For the exterior sphere of our system, we need to obtain the component of the $T$ operator describing a scattering problem in which both the source of the incident wave and the detector  are inside the sphere. Hence, we consider now the sphere subject to the same $\delta$-$\delta'$ potential at the surface, but the source is now inside the body, at its origin. Therefore, the general solution for the  radial part of a field mode is
  \begin{equation}\label{solutions_source_inside}
\rho_\ell(x)=\left\{ 
\begin{array}{cc}
\tilde a_\ell\rho_\ell^{\text{reg}}(x)+\tilde b_\ell\rho_\ell^{\text{out}}(x)  & x<x_0  \\
B_\ell\rho_\ell^{\text{out}}(x)  & x>x_0  \\
\end{array}
\right..
\end{equation}
The coefficients $\{B_\ell,\tilde a_\ell,\tilde b_\ell\}$ above must satisfy the boundary conditions obtained by plugging Eq.~\eqref{solutions_source_inside} into Eq.~\eqref{eq:matchG},
\begin{eqnarray}
 && B_\ell \left(
  \begin{array}
    {c}\rho_\ell^{\text{out}}(x_0)\\
    { \dot{\rho}_\ell^{\text{out}}(x_0)} \\
  \end{array}
  \right)\nonumber \\
  &=&\left(
  \begin{array}{cc}
    \alpha  & 0 \\
    \widetilde{\beta} & {\alpha^{-1} } \\
  \end{array}
  \right)\left(
  \begin{array}{c}
    \tilde a_\ell \rho_\ell^{\text{reg}}(x_0)+\tilde b_\ell \rho_\ell^{\text{out}}(x_0) \\
    {\tilde a_\ell \dot{\rho}_\ell^{\text{reg}}(x_0)+\tilde b_\ell \dot{\rho}_\ell^{\text{out}}(x_0)} \\
  \end{array}
  \right).\label{bc_inner_scattering}
\end{eqnarray}
 As in the previous case the desired component of the $T$ operator is given by the ratio of the reflected flux to the emitted wave, but this time inside the sphere,
  $$\widetilde{\mathbb T}^\ell=-\frac{\tilde a_\ell}{\tilde b_\ell}.$$ 
 The latter can be easily obtained from the Eq.~\eqref{bc_inner_scattering}:
 \begin{eqnarray}
    \widetilde{\mathbb T}^\ell(\omega)&=&\frac{h_\ell^{(1)}(\omega x_0)}{\Lambda(\omega)}\Big\{\big[\ell(\alpha^2-1)-x_0\alpha\tilde\beta\big]h_\ell^{(1)}(\omega x_0)\nonumber \\
   &&-(\alpha^2-1)\omega x_0\,h_{\ell+1}^{(1)}(\omega x_0)\Big\}.\label{tildeT_i}
  \end{eqnarray}

\paragraph{On the relation between $\mathbb{T}$ and $\widetilde{\mathbb{T}}$.} If we compare the numerators in Eqs.~\eqref{T_i} and \eqref{tildeT_i} we can see that the are related by exchanging $j_\ell(\omega x_0)\leftrightarrow h^{(1)}_\ell(\omega x_0)$. The same property does not hold for the  components  $\mathbb T^\ell(\omega)$ and $~\widetilde{\mathbb T}^\ell(\omega)$. This reciprocity corresponds to exchanging the incident and reflected wave. But as we have noted, the interior and exterior sides of the sphere are different so we have also to exchange them. In this sense, taking into account that the inverse of the matching condition matrix appearing in \eqref{bc_outer_scattering} and \eqref{bc_inner_scattering}
\begin{equation}
\left(
\begin{array}{cc}
\alpha \  & 0 \\
\widetilde{\beta} \ & {\alpha^{-1}} \\
\end{array}
\right)^{-1}=\left(
\begin{array}{cc}
\alpha^{-1} \  & 0 \\
-\widetilde{\beta}\ & {\alpha } \\
\end{array}
\right)
\end{equation}
is reached with the coupling transformation $\{\lambda_0,\lambda_1\}\to \{-\lambda_0,-\lambda_1\}$ we conclude that
 \begin{equation}\label{sym-transf}
\widetilde{\mathbb T}^\ell(\omega;x_0,\lambda_0,\lambda_1)={\mathbb T}^\ell(\omega;x_0,-\lambda_0,-\lambda_1;j_\ell\leftrightarrow h^{(1)}_\ell).
\end{equation}
This result is quite surprising when we compare with the one dimensional case that enables to mimic two dimensional plates as was shown in Ref. \cite{munoz2015delta}. For the one dimensional case and the potential $V_{1D}=w_0\delta(x)+2w_1\delta'(x)$ the role played by ${\mathbb T}^\ell$ and $\widetilde{\mathbb T}^\ell$ in our case is played by the reflection amplitudes (see Ref. \cite{munoz2015delta}):
\begin{equation}
r_R=\frac{-\omega\, w_1-i w_0}{2\,\omega\,(w_1^2+1)+iw_0},\quad r_L=\frac{\omega\, w_1-i w_0}{2\,\omega\,(w_1^2+1)+iw_0}.
\end{equation}

In this case it is straightforward to notice that the transformation
\begin{equation}
(w_0,w_1)\mapsto (w_0,-w_1)
\end{equation}
acting on the reflection amplitudes enables us to obtain the analogue of the three dimensional case, i. e.:
\begin{equation}
r_R(\omega;w_0,-w_1)=r_L(\omega;w_0,w_1).
\end{equation}
Hence, meanwhile the symmetry between reflection amplitudes in the one dimensional $\delta\text{-}\delta'$ potential only requires the change of sign of the $\delta'$ coupling, for the spherical three dimensional case it is necessary, in addition, a change of sign of the Dirac $\delta$ coupling as it is shown in \eqref{sym-transf}. This additional requirement implies that the Dirac $\delta$ potentials change from being a potential well/barrier to a barrier/well.  
%
  \section{Analytic expression and numerical results for the Casimir interaction energy}
  %
  In order to use Eq.~\eqref{eq:EnergyTGTG} we remind the reader that in our case object $1$ refers to the interior sphere and therefore the scattering is produced outside. This is described by $\mathbb T_1^\ell$ as given in Eq.~\eqref{T_i} setting $x_0=x_1$, and $(\alpha,\tilde\beta)=(\alpha_1,\tilde\beta_1)$. On the other hand, the waves reaching object $2$ are scattered from the inside and therefore it corresponds to $\widetilde {\mathbb T}_2^\ell$ as in Eq.~\eqref{tildeT_i} with  $x_0=x_2$, and $(\alpha,\tilde\beta)=(\alpha_2,\tilde\beta_2)$. In addition, we need to obtain the expressions for the $T$ operators for imaginary frequencies in order to use the TGTG formula in its euclidean version, where the $T$ operators are Hermitian and oscillatory behavior in the integrals is avoided. Therefore, we define $\omega= i\chi$ with $\chi>0$. The Bessel functions with imaginary arguments can be written in terms of the modified Bessel functions of the first and second kind \cite{olver2010nist},
  \begin{eqnarray*}
 &&   j_\ell(i\chi x)=i^\ell\sqrt{\frac{\pi}{2\chi x}}I_{\ell+1/2}(\chi x)\\
 && h^{(1)}_\ell(i\chi x)=-i^{-\ell}\sqrt{\frac{2}{\pi\chi x}}K_{\ell+1/2}(\chi x).
  \end{eqnarray*}
Taking into account the equations above and Eqs. \eqref{T_i} and \eqref{tildeT_i} the  euclidean rotated components of the required $T$ operators become
\begin{widetext}
\begin{equation}\label{eq:Tcomponents}
\begin{array}{r@{}l}
  \mathbb T_1^\ell(i\chi)\,&= \rm C\,\,\dfrac{ {\it I}_\nu\left({\it y}_1\right)\Big[{\it I}_\nu\left({\it y}_1\right) \left(\ell(\alpha_1^2-1)-\alpha_1{\it x}_1\widetilde{\beta }_1\right)+\left(\alpha_1^2-1\right) {\it y}_1{\it I}_{\nu+1}\left({\it y}_1\right)\Big]}{\Xi({\it y}_1)},\\
\tilde {\mathbb T}_2^\ell(i\chi)\,&=\rm C^{-1}\,\,\dfrac{{\it K}_\nu\left({\it y}_2\right)\Big[{\it K}_\nu\left({\it y}_2\right) \left(\ell(\alpha_{2}^2-1)-\alpha_{2} {\it x}_2 \widetilde{\beta }_2\right)-\left(\alpha_{2}^2-1\right) {\it y}_2 {\it K}_{\nu+1}\left({\it y}_2\right)\Big]}{\Xi({\it y}_2)},\\
\end{array}
\end{equation}
\begin{equation*}
\Xi(y_{i})\equiv I_\nu\left(y_{i}\right)\Big[K_\nu\left(y_{i}\right)\left(\ell\alpha_{i}^2-\ell-\alpha_{i}x_{i}\widetilde{\beta }_{i}\right)
 -\alpha_{i}^2y_{i} K_{\nu+1}\left(y_{i}\right)\Big]-y_{i}I_{\nu+1}\left(y_{i}\right)K_\nu\left(y_{i}\right),
\end{equation*}
\end{widetext}
where $\rm C \equiv (-1)^\ell(\pi/2)$, $\nu\equiv\ell+1/2$, and ${\it y}_{\,{i}}\equiv\chi {\it x}_{\,{i}}$, for $ i=1,2$.
Before computing the quantum vacuum energy, as a consistency test, it is straightforward to observe that by turning off the $\delta'$ term $\lambda_{1,{i}}=0$ ($\alpha_{i}=1,\,\, \beta_{i}=\lambda_{0,{i}}$), equations in \eqref{eq:Tcomponents} become
\begin{equation}
\begin{array}{r@{}l}\label{tdleta}
\mathbb T_1^\ell(i\chi)&\,= \rm C\,\dfrac{\lambda_{0,1}{\it x}_1{\it I}_\nu^2\left({\it y}_1\right)}{[1+\lambda_{0,1}{\it x}_1{\it K}_{\nu}({\it y}_1){\it I}_{\nu}({\it y}_1)]},\\
\tilde {\mathbb T}_2^\ell(i\chi)&\,=\rm C^{-1}\,\dfrac{\lambda_{0,2}{\it x}_2{\it K}_\nu^2\left({\it y}_2\right)}{[1+\lambda_{0,2}{\it x}_2{\it K}_{\nu}({\it y}_2){\it I}_{\nu}({\it y}_2)]}.
\end{array}
\end{equation}

\noindent These expressions are in agreement with the ones for two concentric spheres having delta potentials on their surfaces calculated in \cite{teo2012mode} and \cite{milton2008multiple}. See also Ref. \cite{beauregard2015casimir} where they calculate the same $\mathbb T$ through the phase shift.

In this sense, the TGTG formula for the interaction energy when both spheres share center can be written as 
\begin{equation}\label{eq:EnergyTT}
\!\!\!E_{\text{C}}=\frac{1}{2\pi}\sum_{\ell=0}^{\infty}(2\ell+1)\int_0^{\infty} d\chi \hspace{0mm}\ln  \Big[1-\mathbb T_1^\ell(i\chi)\widetilde {\mathbb T}_2^\ell(i\chi)\Big].
\end{equation}
The expressions in Eqs. \eqref{eq:Tcomponents} and \eqref{eq:EnergyTT} enable us to obtain numerical results for the quantum vacuum interaction energy between the two concentric spheres. Regarding the presentation of the numerical plots of the quantum vacuum energy we consider different possible scenarios by changing the couplings in the potential. In all the cases shown bellow, we take the coefficient of the $\delta$ term to be positive and allow the coefficient of the $\delta'$ to change sign. As it has been seen along the paper, the presence of the $\delta'$ term makes the potential on the spheres behave differently when the scattering is produced from the inside or from the outside of the body. 

First we consider the couplings to be equal in both spheres, such that $\lambda_{0,1}=\lambda_{0,2}=\lambda_0$ and $ \lambda_{1,1}= \lambda_{1,2}=\lambda_1$. We show the results in Fig.~\ref{fig:2} for two different values of the radii. In the plot on the left we have used $x_1=1$  and $x_2=2$, and $x_1=1.8$  and $x_2=2$ in the one on the right. The color gradient denotes changes on the energy value.
We observe that in both plots there are regions in the space of couplings where the vacuum energy takes positive, negative and zero values. When $\lambda_1=0$ we recover the case of the interaction between two $\delta$ (semitransparent) spheres that is known to be negative. The potentials in both spheres have the same sign. When $\lambda_0=0$, and since we have set the couplings of the $\delta'$ term equal, the sign of the potentials on each sphere is determined by the sign of the $\delta'$ that we know behaves differently from the inside and outside. Consequently, the interaction energy becomes positive. When both terms are present in the potential, one of them is dominant over the other. As $\lambda_{0}$ increases, higher absolute values of $\lambda_1$ are needed to obtain a positive energy. This pattern holds for both plots, although the numerical values depend on the radii.

We also observe in Fig.~\ref{fig:2} that  the $\delta'$ contribution is not symmetric under $\lambda_1\to -\lambda_1$. In the right graph, where  the radii of the spheres do not differ much from each other {(and therefore the situation approaches the parallel plates configuration when the radii are large enough}) we see that if both couplings of the $\delta'$ are positive $\lambda_1>0$,  the Casimir energy shows almost the same pattern as in the case when both are negative $ \lambda_1<0$. This symmetry fades out as the difference between the values of the radii increases (see plot on the left); that means, when the inner sphere becomes comparatively smaller than the outer one.
Next we turn off the delta interaction in the potential by doing $\lambda_{0,1}=\lambda_{0,2}=0$, so that we are left with concentric spheres defined by a $\delta'$ potential alone on their surfaces. Results are shown in the right graph in Fig. \ref{fig:3} for $x_1=1$ and $x_2=2$. We observe that when the couplings have the same sign the interaction energy is positive (as we mentioned above), while it becomes negative if the couplings have different sign. We compare this result with the equivalent one from a plane geometry showed on the left of Fig. \ref{fig:3}, where the same pattern is obtained. For both geometries the results agree with the change in sign that the $\delta'$ introduces when it is approached from inside or outside, or equivalently for planar geometry, from one side or another. We furthermore observe again how the spherical geometry introduces an asymmetry on the values of the positive and negative energies compared with parallel plates.\\

We test the numerical results by making the radii of the spheres large while keeping a small constant the difference between them so that we can compare with the parallel plates geometry. The plots are presented in Fig.~\ref{fig:1}. We see a tendency to recover  the behaviour of the Casimir energy for planar geometry studied in Ref.~\cite{munoz2015delta}. The plots show the interaction Casimir energy for different values of the couplings when these are the same in both bodies.

\begin{widetext}
  
\begin{figure}[h]
  \centering
  \includegraphics[width=0.475\textwidth]{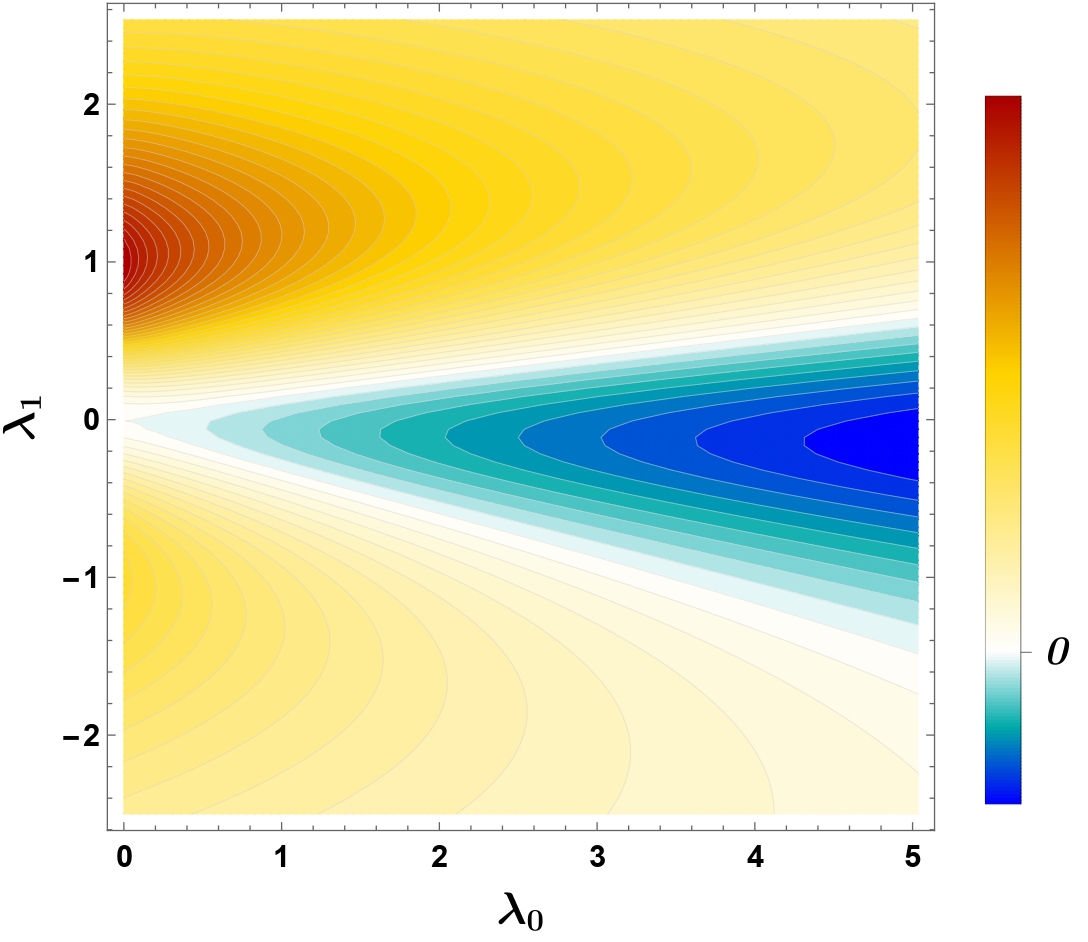}%
  \quad
  \includegraphics[width=0.475\textwidth]{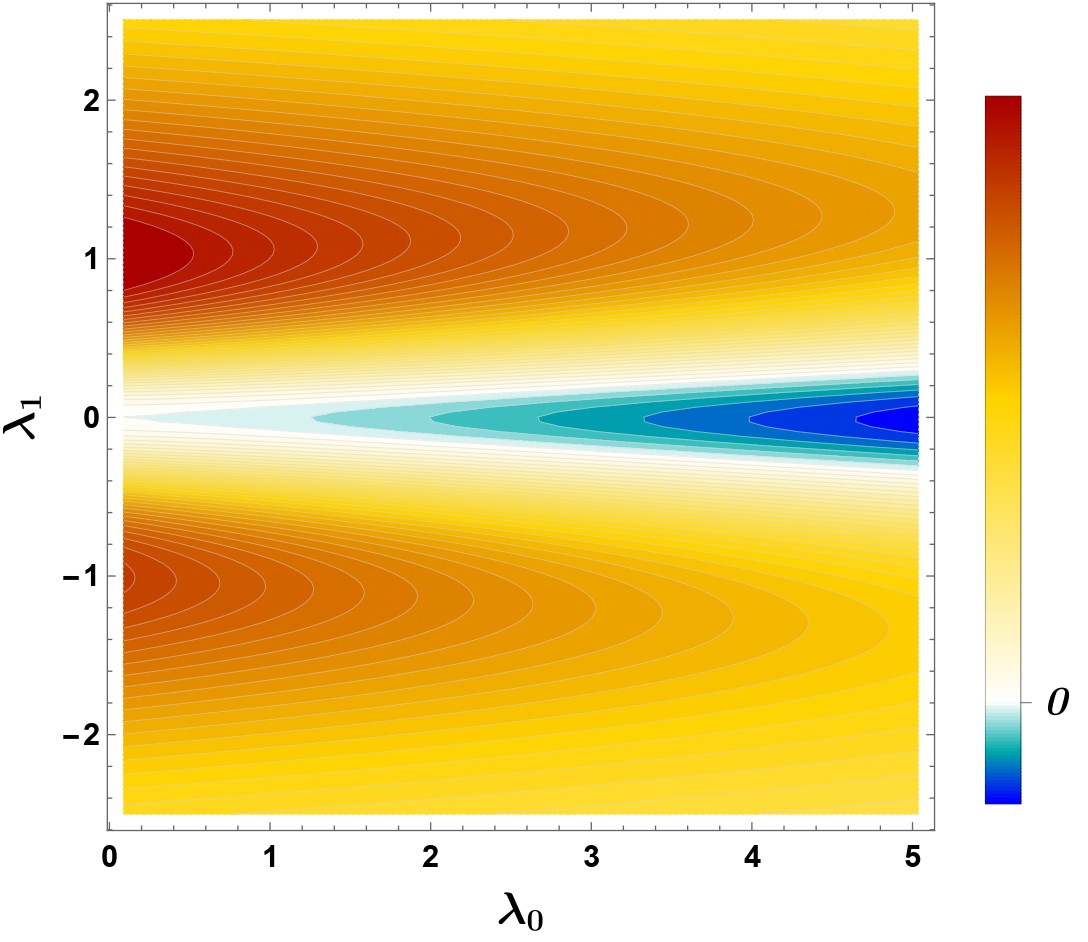} %
  \caption{The quantum vacuum interaction energy obtained from Eq.~\eqref{eq:EnergyTT} when $\lambda_{0,1}=\lambda_{0,2}=\lambda_0$ and $ \lambda_{1,1}= \lambda_{1,2}=\lambda_1$. In the LEFT plot: radii $x_1=1$ and $x_2=2$. In the RIGHT plot: radii $x_1=1.8$ and $x_2=2$}%
  \label{fig:2}%
\end{figure}

\begin{figure}
  \centering
  \includegraphics[width=0.475\textwidth]{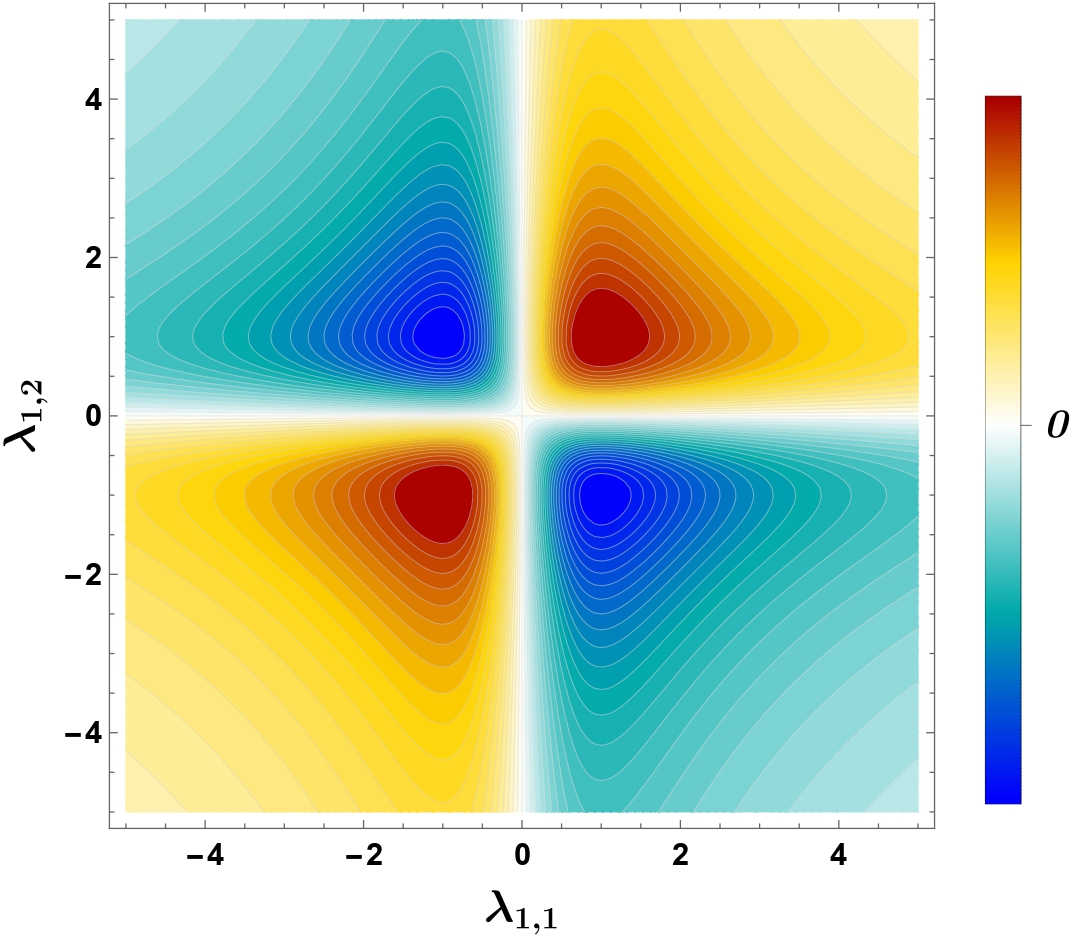}
  \quad
  \includegraphics[width=0.475\textwidth]{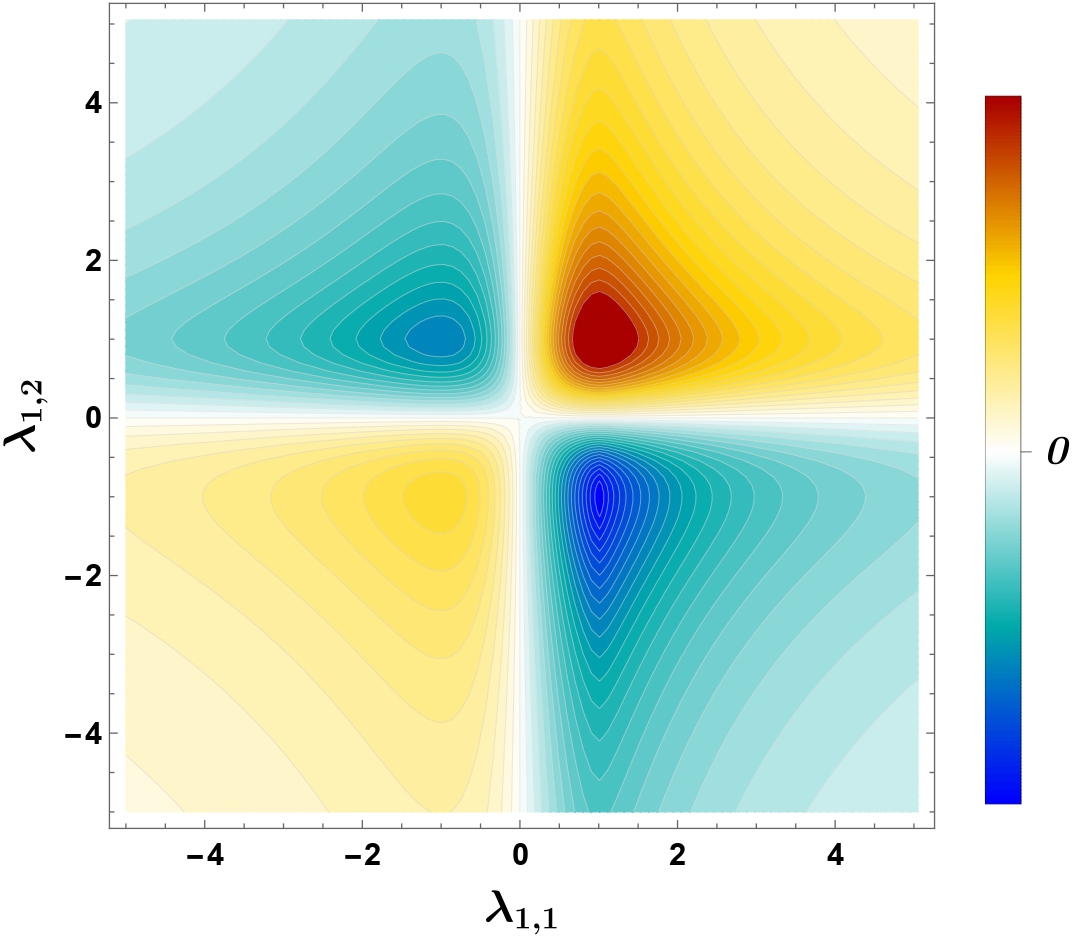} %
  \caption{Comparison between the quantum vacuum interaction energy of two $\delta$-$\delta'$ plane parallel plates and two concentric $\delta$-$\delta'$ spheres with $\lambda_{0,1}=\lambda_{0,2}=0$. The LEFT plot: two plates separated unit distance. RIGHT plot: spherical shells with $x_1=1$ and $x_2=2$}%
  \label{fig:3}%
\end{figure}

\begin{figure}
  \centering
  \includegraphics[width=0.475\textwidth]{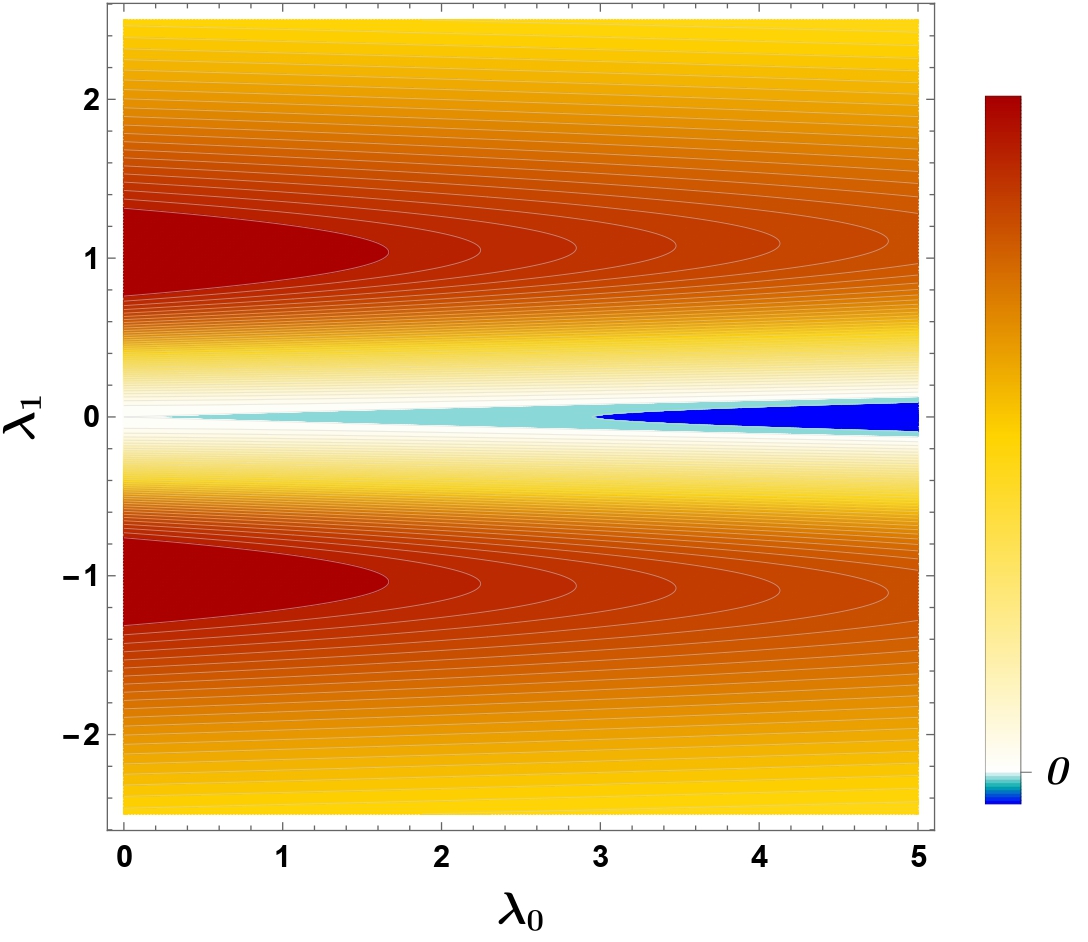} %
  \quad
  \includegraphics[width=0.475\textwidth]{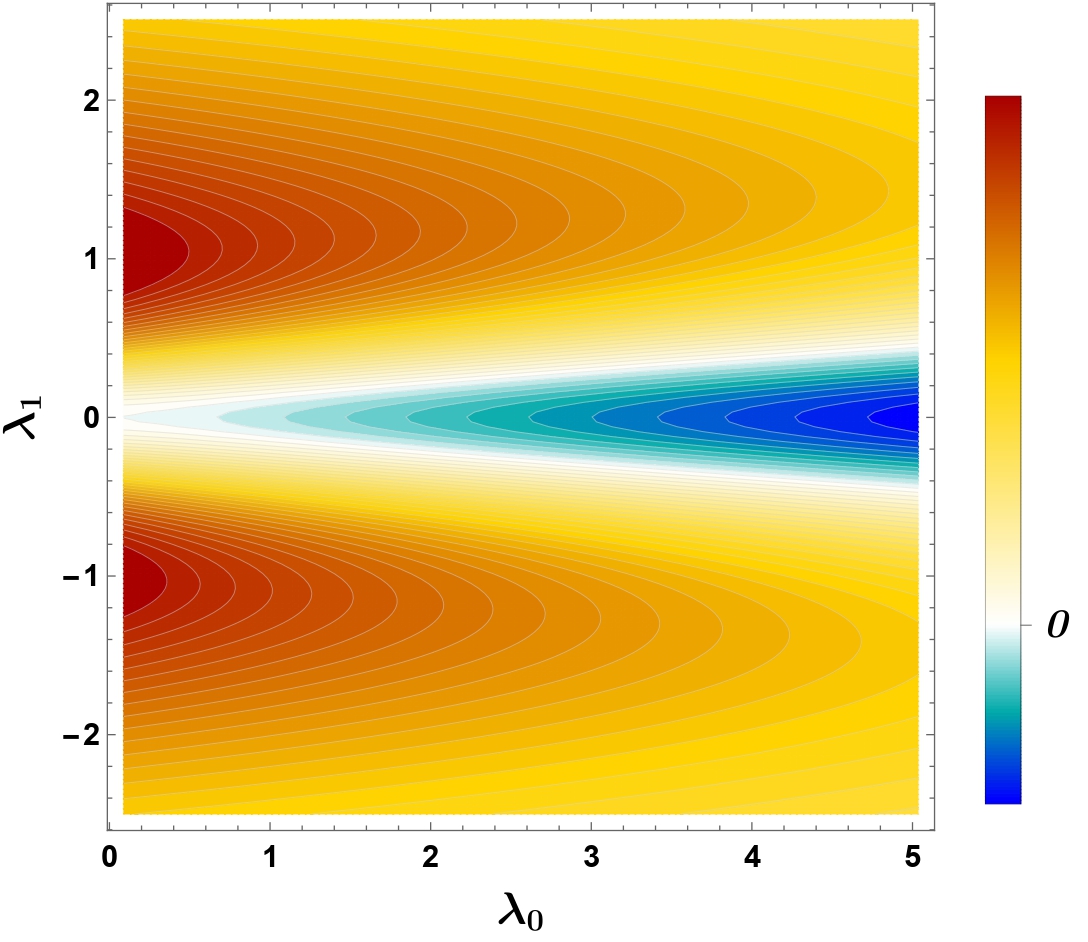} %
  \caption{Effect of the distance on the quantum vacuum interaction energy of two $\delta$-$\delta'$ spheres for $\lambda_{0,1}=\lambda_{0,2}=\lambda_0$ and $ \lambda_{1,1}= \lambda_{1,2}=\lambda_1$. LEFT plot: plates separated 0.1 units of distance. RIGHT plot: radii $x_1=10$ and $x_2=10.1$}%
  \label{fig:1}%
\end{figure}

\begin{figure}
  \centering
  \includegraphics[width=0.475\textwidth]{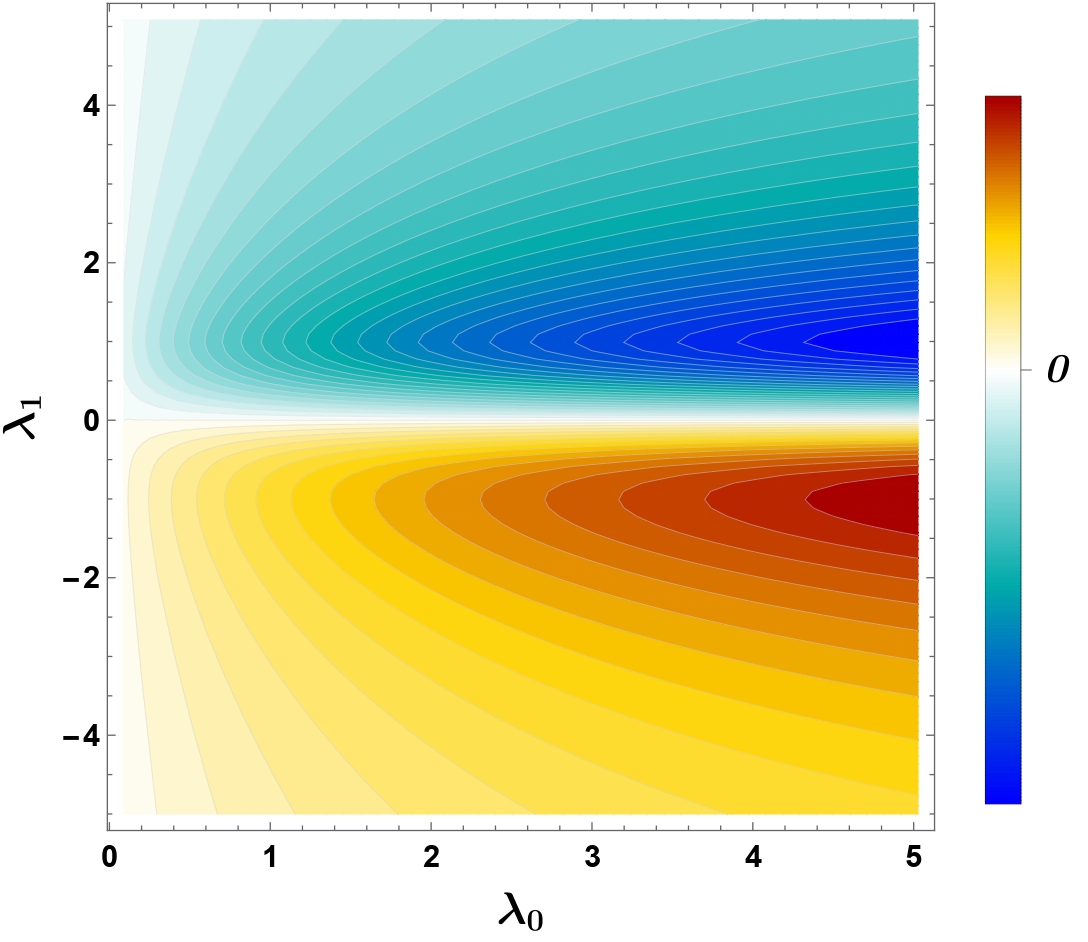} %
  \qquad
  \includegraphics[width=0.475\textwidth]{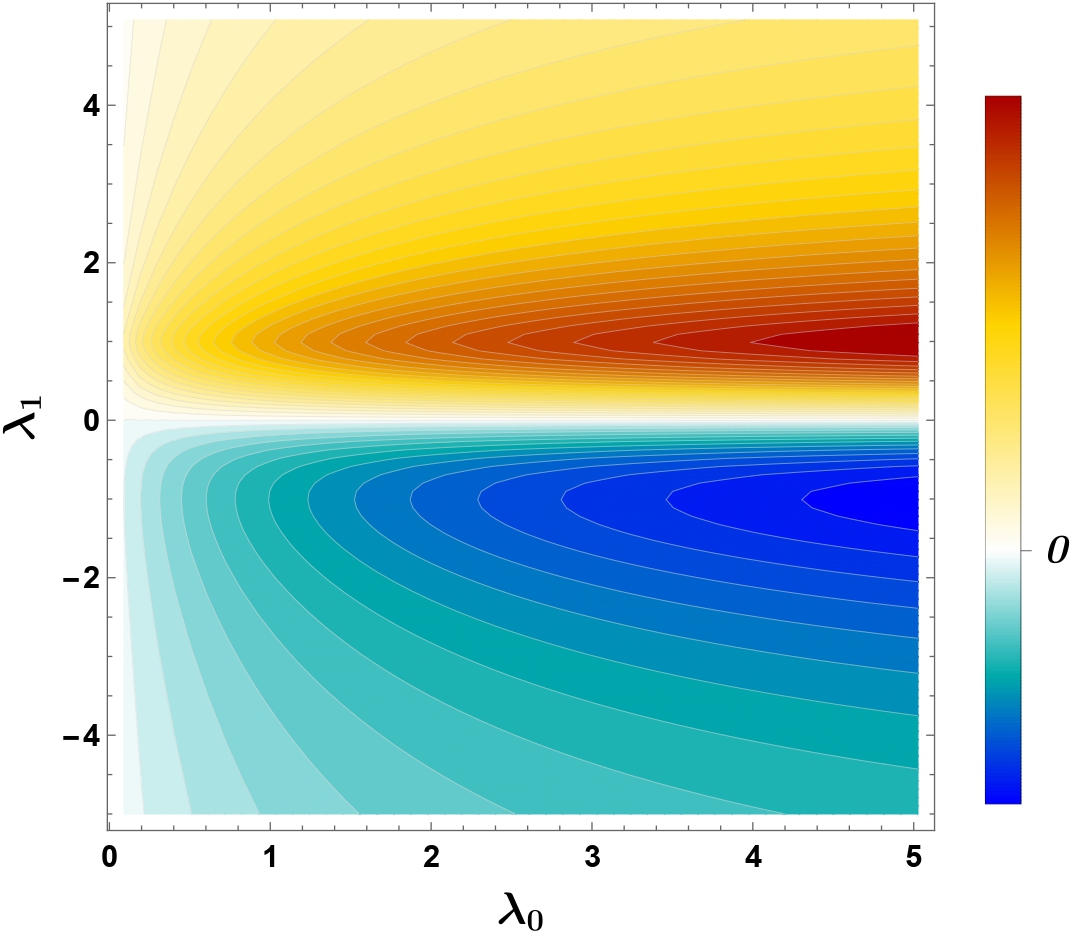} %
  \caption{Radii $x_1=1$ and $x_2=2$. LEFT plot: $\delta$ \textit{vs} $\delta'$: $\lambda_{1,1}=\lambda_{0,2}=0$ . RIGHT plot: $\delta'$ \textit{vs} $\delta$: $\lambda_{0,1}=\lambda_{1,2}=0$.}%
  \label{fig:4}%
\end{figure}

\end{widetext}

The plot on the left shows the result for parallel plates while the one on the right is generated from concentric spheres with large radii keeping values with small difference between them. It can be seen that in this situation there is a tendency to recover the behaviour of the quantum vacuum interaction energy between two plates as the values of $x_1$ and $x_2$ increase keeping constant the distance between them.

Finally, in Fig.~\ref{fig:4} we consider the case in which one sphere is defined by a $\delta$ and the other one by a $\delta'$ interaction. As expected, the sign of the interaction energy changes from one setup to the other illustrating the influence of having the $\delta'$ hit from the interior sphere or the exterior one.

We wrap up this section stressing a common feature in the plots showed.
We observe maximum absolute values of  $E_{\text{C}}$ when $|\lambda_{1}|=1$. In this case the matching conditions \eqref{eq:matchG} are ill defined and they transform into Robin or Dirichlet boundary conditions \cite{munoz2015delta, romaniega2020approximation}.

Again, the $\delta'$ term makes the matching condition different form one side of the body than from the other,
\begin{equation}\label{eq:RobinDirichlet}
  \begin{aligned}
     \rho_{\ell}(x_0^-)&=0, \quad \dot{\rho}_{\ell}(x_0^+)=-D\rho_{\ell}(x_0^+) \quad\text{if}\quad \lambda_1= 1,\\
  \dot{\rho}_{\ell}(x_0^-)&=D\rho_{\ell}(x_0^-), \quad\rho_{\ell}(x_0^+)=0\quad\text{if}\quad \lambda_1=-1,
\end{aligned}
\end{equation}
where $D=4/(\lambda_0-4x_0)$ is a constant on the sphere.
For example, in Fig. \ref{fig:3} we see that higher values of the positive energy are achieved 
for $\lambda_{1,1}=\lambda_{1,2}=1$ (Robin \textit{vs} Dirichlet) rather than for $\lambda_{1,1}=\lambda_{1,2}=-1$ (Dirichlet \textit{vs} Robin). For negative energies 
$|E_{\text{C}}|$ reaches higher values
for $\lambda_{1,1}=-\lambda_{1,2}=1$ (Robin \textit{vs} Robin) than for $\lambda_{1,1}=-\lambda_{1,2}=-1$ (Dirichlet \textit{vs} Dirichlet). In Figs.~\ref{fig:2} and \ref{fig:1} the two local maximum values of $E_{\text{C}}$
are reached for $|\lambda_1|=1$ with $\lambda_0=0$. The same holds in Fig.~\ref{fig:4}, but  $|E_{\text{C}}|$ grows with $\lambda_{0}$ in the range considered. Modeling the spheres in this way, we can study cases where one of the spheres behaves purely electric, by imposing Dirichlet boundary conditions that correspond to TE modes, and the other purely magnetic, by imposing Robin boundary conditions that correspond to TM modes, or any other possible combination.

\section{Conclusions}
%
We have computed the quantum vacuum interaction energy between two concentric spheres mimicked by spherically symmetric $\delta$-$\delta'$ potentials. We have used the TGTG formula stressing the difference between the two T operators that enter the system denoted by $\mathbb{T}$ and $\widetilde{\mathbb{T}}$.\\

The analytical expressions given in Eqs. \eqref{T_i} and \eqref{tildeT_i}
allowed us to study in detail the physical interpretation of the socalled $\widetilde{\mathbb{T}}$ operator in terms of a nonstandard scattering problem where the source of incident probability flux is placed in the centre of the sphere instead of being placed at infinity as it happens in most standard scattering problems.\\

In addition, the analytical results from Eqs. \eqref{T_i} and \eqref{tildeT_i} enables us to relate the $\widetilde{\mathbb{T}}$ operator with the more common $\mathbb{T}$ operator by means of the symmetry transformation given in Eq.~\eqref{sym-transf}. The mentioned transformation requires the change in sign of the coupling of the $\delta$ potential unlike it happens for the same potential in the one-dimensional case. \\

By using Eqs. \eqref{eq:Tcomponents} and \eqref{eq:EnergyTT} we have been able to obtain numerical results for the quantum vacuum interaction energy of two concentric $\delta$-$\delta'$ spheres as a function of the four free parameters entering in the potential. As a result, it can be seen, in Figs.~\ref{fig:2}-\ref{fig:4}, that the quantum vacuum interaction energy has not a well defined sign as a function of the parameters $\{\lambda_{0,i},\lambda_{1,i}\}_{i=1,2}$.
The positive energy values are clearly due to the presence of the $\delta'$ term since we have considered positive contributions of the $\delta$ potentials $\lambda_{0,i}>0$. This is due to the fact that the $\delta'$ term behaves differently on one side of the sphere and the other, causing a change of sign and affecting the boundary conditions on the sphere.\\

We observe maximum values of the quantum vacuum interaction energy for couplings of the $\delta'$ equal to $1$ or $-1$. The potential reported could be equivalent to consider Robin boundary conditions. Moreover, for certain values of the couplings we can achieve purely Dirichelt or Neumann boundary conditions. We have shown that our result can also be extrapolated with success to limiting cases as parallel plates with $\delta$-$\delta'$ potential or concentric $\delta$ spheres.
 
\section*{Acknowledgments}

ICP would like to thank funding from the grant DGA: E21-17R. JMMC and CR are grateful to the Spanish Government-MINECO (MTM2014-57129-C2-1-P) and the Junta de Castilla y Le\'on (BU229P18, VA137G18 and VA057U16) for the financial support received. CR is grateful to MINECO for the FPU fellowship programme (FPU17/01475).
%
%

%

\begin{thebibliography}{10}

\bibitem{klimchitskaya2009casimir}
G. L. Klimchitskaya, U. Mohideen, and V. M. Mostepanenko,
\newblock {Rev. Mod. Phys.} \textbf{81}, 1827 (2009).

\bibitem{munday2009measured}
J.~N. Munday, F.~Capasso, and V.~A. Parsegian,
\newblock {Nature} \textbf{457}, 170 (2009).

\bibitem{chan2001quantum}
H.~B. Chan, V.~A. Aksyuk, R.~N. Kleiman, D.~J. Bishop, and F.~Capasso,
\newblock { Science}  \textbf{291}, 1941 (2001).

\bibitem{bordag2009advances}
M.~Bordag, G.~L. Klimchitskaya, U.~Mohideen, and V.~M. Mostepanenko,
\newblock {\em Advances in the Casimir Effect}
\newblock (Oxford
University Press, New York, 2009).

\bibitem{dalvit2011casimir}
D.~Dalvit, P.~Milonni, D.~Roberts, and F.~Da~Rosa,
\newblock {\em Casimir Physics}
\newblock (Springer, New York, 2011).

\bibitem{milton2004casimir}
K.~A. Milton,
\newblock { J. Phys. A Math. Gen.} \textbf{37}, R209 (2004).

\bibitem{candelas1982vacuum}
P.~Candelas,
\newblock { Ann. Phys.} \textbf{143}, 241 (1982).

\bibitem{candelas1986vacuum}
P.~Candelas,
\newblock { Ann. Phys.} \textbf{167}, 257 (1986).

\bibitem{cavero2006local}
I.~Cavero-Pel{\'a}ez, K.~A. Milton, and J.~Wagner,
\newblock { Phys. Rev. D} \textbf{73}, 085004 (2006).

\bibitem{casimir1948attraction}
H.~B.~G. Casimir,
\newblock{Proc. Kon. Ned. Akad. Wet.} \textbf{51}, 793 (1948).

\bibitem{mazzitelli2003casimir}
F.~D. Mazzitelli, M.~J. S{\'a}nchez, N.~N. Scoccola, and J.~von Stecher,
\newblock { Phys. Rev. A} \textbf{67}, 013807 (2003).

\bibitem{brevik2002casimir}
I.~Brevik, J.~B. Aarseth, and J.~S. H{\o}ye,
\newblock { Phys. Rev. E} \textbf{66}, 026119 (2002).

\bibitem{saharian2001scalar}
A.~A. Saharian,
\newblock {Phys. Rev. D} \textbf{63}, 125007 (2001).

\bibitem{teo2012mode}
L.~P. Teo,
\newblock {Int. J. Mod. Phys. A} \textbf{27}, 1230021 (2012).

\bibitem{marachevsky2001casimir}
V.~N. Marachevsky,
\newblock { Phys. Scr.} \textbf{64},  (2001).

\bibitem{milton2011casimir}
K.~A. Milton, E.~K. Abalo, P.~Parashar, N.~Pourtolami, I.~Brevik, and S.~{\AA}.
  Ellingsen,
\newblock { Phys. Rev. A} \textbf{83}, 062507 (2011).

\bibitem{kenneth2006opposites}
O.~Kenneth and I.~Klich,
\newblock { Phys. Rev. Lett.} \textbf{97}, 160401  (2006).

\bibitem{emig2008casimir}
T.~Emig, N.~Graham, R.~L. Jaffe, and M.~Kardar,
\newblock {Phys. Rev. D} \textbf{77}, 025005 (2008).

\bibitem{kenneth2008casimir}
O.~Kenneth and I.~Klich,
\newblock {Phys. Rev. B} \textbf{78}, 014103 (2008).

\bibitem{rahi2009scattering}
S.~J. Rahi, T.~Emig, N.~Graham, R.~L. Jaffe, and M.~Kardar,
\newblock {Phys. Rev. D} \textbf{80}, 085021 (2009).

\bibitem{zaheer2010casimir}
S.~Zaheer, S.~J. Rahi, T.~Emig, and R.~L. Jaffe,
\newblock { Phys. Rev. A} \textbf{82}, 052507 (2010).

\bibitem{rahi2010stable}
S.~J. Rahi and S.~Zaheer,
\newblock {Phys. Rev. Lett.} \textbf{104}, 070405 (2010).

\bibitem{rahi2010constraints}
S.~J. Rahi, M.~Kardar, and T.~Emig,
\newblock { Phys. Rev. Lett.} \textbf{105}, 070404 (2010).

\bibitem{bimonte2016}
G.~Bimonte,
\newblock {Phys. Rev. D} \textbf{94}, 085021 (2016).

\bibitem{bordag2009vacuum}
M. Bordag and V. Nikolaev,
\newblock {J. Phys. A: Math. Theor.} \textbf{42}, 415203 (2009).

\bibitem{beauregard2015casimir}
M.~Beauregard, M.~Bordag, and K.~Kirsten.
\newblock { J. Phys. A: Math. Theor.} \textbf{48}, 095401 (2015).

\bibitem{BordagPRD1996}
M.~Bordag and K.~Kirsten,
\newblock {Phys. Rev. D} \textbf{53},~5753 (1996).

\bibitem{Fulling89}S.~Fulling,
\newblock {\it Aspects of Quantum Field Theory in Curved Spacetime}
\newblock (Cambridge University Press 1989).

\bibitem{CRS20}
C.~Romaniega,
  \newblock {arXiv:} 2008.02031.

\bibitem{lippmann1950variational}
B.~A. Lippmann and J.~Schwinger,
\newblock {Phys. Rev.} \textbf{79}, 469 (1950).

\bibitem{munoz2019hyperspherical}
J.~M. Mu{\~n}oz-Casta{\~n}eda, L.~M. Nieto, and C.~Romaniega,
\newblock {Ann. Phys.} \textbf{400}, 246 (2019).

\bibitem{bordag1999ground}
M.~Bordag, K.~Kirsten, and D.~Vassilevich,
\newblock {Phys. Rev. D} \textbf{59}, 085011 (1999).

\bibitem{barton2004casimir}
G.~Barton,
\newblock {J. Phys. A Math. Gen.} \textbf{37}, 1011 (2004).

\bibitem{cavero2007local}
I.~Cavero-Pel{\'a}ez, K.~A. Milton, and K.~Kirsten,
\newblock {J. Phys. A: Math. Theor.} \textbf{40}, 3607 (2007).

\bibitem{parashar2017electromagnetic}
P.~Parashar, K.~A. Milton, K.~V. Shajesh, and I.~Brevik,
\newblock {Phys. Rev. D} \textbf{96}, 085010 (2017).

\bibitem{bordag2015dirac}
M.~Bordag and J.~M. Munoz-Castaneda,
\newblock {Phys. Rev. D} \textbf{91}, 065027 (2015).

\bibitem{bordag2017casimir}
M.~Bordag and I.~G. Pirozhenko,
\newblock {Phys. Rev. D} \textbf{95}, 056017 (2017).

\bibitem{bordag2014monoatomically}
M.~Bordag,
\newblock {Phys. Rev. D} \textbf{89}, 125015 (2014).

\bibitem{munoz2015delta}
J.~M. Mu{\~n}oz-Casta{\~n}eda and J.~M. Guilarte,
\newblock {Phys. Rev. D} \textbf{91}, 025028 (2015).

\bibitem{bordag2001new}
M.~Bordag, U.~Mohideen, and V.~M. Mostepanenko,
\newblock {Phys. Rep.} \textbf{353}, 1 (2001).

\bibitem{bachas2007comment}
C.~P. Bachas,
\newblock {J. Phys. A} \textbf{40}, 9089 (2007).

\bibitem{kurasov1996distribution}
P.~Kurasov,
\newblock {J. Math. Anal. Appl.} \textbf{201}, 297 (1996).

\bibitem{romaniega2020approximation}
C.~Romaniega, M.~Gadella, R.~M.~Id Betan, and L.~M. Nieto,
\newblock {Eur. Phys. J. Plus} \textbf{135}, 372 (2020).

\bibitem{balian1977electromagnetic}
R.~Balian and B.~Duplantier,
\newblock {Ann. Phys.} \textbf{104}, 300 (1977).

\bibitem{balian1978electromagnetic}
R.~Balian and B.~Duplantier,
\newblock {Ann. Phys.} \textbf{112}, 165 (1978).

\bibitem{graham2002calculating}
N.~Graham, R.~L. Jaffe, V.~Khemani, M.~Quandt, M.~Scandurra, and H.~Weigel,
\newblock {Nucl. Phys. B} \textbf{645}, 49 (2002).

\bibitem{jaffe2005casimir}
R. L.~Jaffe,
\newblock {Phys. Rev. D} \textbf{72}, 021301 (2005).

\bibitem{emig2008fluctuation}
T.~Emig,
\newblock {J. Stat. Mech. Theory Exp.} \textbf{2008}, P04007 (2008).

\bibitem{canaguier2009casimir}
A.~Canaguier-Durand, P.~A.~MaiaNeto, I.~Cavero-Pelaez, A.~Lambrecht, and
  S.~Reynaud,
\newblock {Phys. Rev. Lett.} \textbf{102}, 230404 (2009).

\bibitem{taylor2006scattering}
J.~R. Taylor,
\newblock {\em Scattering Theory: The Quantum Theory of Nonrelativistic
  Collisions}
\newblock (Courier Corporation, 2006).

\bibitem{olver2010nist}
F.~W.~J. Olver, D. W. Lozier, R. F. Boisvert, and C. W. Clark,
\newblock {\em NIST Handbook of Mathematical Functions}
\newblock (Cambridge University Press, New York, 2010).

\bibitem{milton2008multiple}
K.~A. Milton and J.~Wagner,
\newblock {J. Phys. A: Math. Theor.} \textbf{41}, 155402 (2008).


\end{thebibliography}
\end{document}